\documentclass[doublecol,linenumbers]{epl2}
\usepackage{amsmath}
\usepackage{graphicx}

\title{Single shot imaging of trapped Fermi gas }

\author{Mariusz Gajda, Jan Mostowski, Tomasz Sowi\'nski, and Magdalena Za\l uska-Kotur }
\shortauthor{M. Gajda \etal}

\institute{
  Institute of Physics, Polish Academy of Sciences, Al. Lotnikow 32/46, PL-02668 Warsaw, Poland
  }

\abstract{
Recently developed techniques allow for simultaneous measurements of the
positions of all ultra cold atoms in a trap with high  resolution. Each such single shot
experiment detects one element of the quantum ensemble formed by the cloud
of atoms. Repeated single shot measurements can be used to determine all
correlations between particle positions as opposed to standard
measurements that determine particle density or two-particle correlations
only. In this paper we discuss the possible outcomes of such single shot
measurements in case of cloud of ultra-cold non-interacting Fermi atoms. We show
that the Pauli exclusion principle alone leads to correlations between particle
positions that originate from unexpected spatial structures formed by the
atoms.
}

\begin{document}

\maketitle

\section{Introduction}
Tremendous progress in experimental techniques of preparing, manipulating and probing ultra-cold gases have opened new possibilities of optical methods of monitoring atomic systems.  Atomic fluorescence microscopes with resolution in the range of hundreds of nanometers became accessible\cite{Bakr1,Bakr2,Sherson,Cheuk,Parsons,Haller,Edge}. The microscopes allow for observation of both boson and fermion atoms with resolution comparable to the optical wavelength. Single shot pictures of such systems correspond to a  single realization of the $N$-body probability density as opposed to a one-particle probability distribution. Difference  between the two is tremendous, they differ by  $N$ body correlations.  The seminal work of \cite{Javanainen} shows how interference fringes, visible in a simultaneous single shot picture of $N$ atoms, arise in the course of measurement. No fringes  are observed in a single particle detection instead. In a similar way the solitons emerge in a process of detection of $N$-particles prepared in a type II excited state of  a 1D system of bosons interacting via short-range potential described by the Lieb-Linger model\cite{Sacha}. Single shot   time-dependent simulations of many-body dynamics showing appearance of fluctuating vortices and center-of-mass fluctuations of attractive BEC have been reported recently \cite{Kasevich}.

$N$-body system is not a simple $N$-fold sum of systems of one particle. This is because of correlations between
particles resulting from their mutual interactions. In quantum systems correlations can be imposed not
only by interactions, but also by the quantum statistics.

Quantum Mechanics  gives a completely different meaning to the classical concept of identical objects \cite{Weinberg}. Quantum identical particles are identical not only because they share the same mass, spin, charge, etc., but also because they cannot be identified by tracing their history. Here we show yet another consequence of quantum indistinguishability. We show that identical fermions confined by an external trapping potential arrange themselves in spectacular geometric structures  even if no mutual interaction is present. This is because the indistinguishability of fermions, formulated in the language of the Pauli exclusion principle, prevents any two fermions from being at the same location. These unexplored geometric structures, Pauli crystals, emerge repeatedly in single shot pictures of the many-body system.

\section{Pauli crystals}

Here we study on a theoretical ground a manifestation of the quantum statistics, namely a high order geometric
correlations in a small system of ultra cold spin polarized fermions confined in space by an external binding potential.  To this end  we generate a single shot picture of this noninteracting system.  We limit our attention to the many-body ground state.  Atoms are attracted towards the trap minimum, but on the other hand, the Pauli exclusion principle does not allow any two fermions to be at the same position. These two competing effects  might, in principle,  lead to a kind of equilibrium.

We limit our attention to  a simple generic example of particles bound by a harmonic potential in two dimensions and  frequency  $\omega_x=\omega_y=\omega$. One-particle states are the standard harmonic oscillator wave functions:
\begin{equation}
\psi_{nm}(x,y)={\cal N}_{nm}\mathrm{e}^{-(x^2+y^2)/2}{\cal H}_n(x){\cal H}_m(y),
\end{equation}
where ${\cal N}_{nm}=( 2^{n+m}n!m!\sqrt{\pi} )^{-1/2}$ is the norm, and ${\cal H}_n(z)$ is the $n$-th Hermite polynomial. The positions $x$ and $y$ are expressed in the normal harmonic oscillator units, i.e. the unit of length being $a=\sqrt{\hbar/M\omega}$, where $M$ is the mass of the particle. Quantum numbers $n$ and $m$ enumerate excitations in $x$ and $y$ direction respectively. We consider an isotropic trap, therefore all states with the same total number of excitations, $n+m$, are degenerated. These states have energy $E_{nm}=\hbar\omega(n+m+1)$, all states of the same energy form an energy shell.

The ground state of a non-interacting $N$-body system is very simple,  every particle occupies a different one-particle state. As a result the $N$ lowest energy states, up to the Fermi energy are occupied. For $N=1,3,6,10,15$  the ground state is uniquely defined because all states at or below the Fermi level are occupied and states above the Fermi level remain not occupied.  The many-body ground state is degenerated whenever the total number of particles does not coincide with the degeneracy of the energy shells.

The many-body wave function is simply the Slater determinant of the occupied one-particle orbitals:  $\Psi(\boldsymbol{r}_1,\ldots,\boldsymbol{r}_N)=\sqrt{\frac{1}{N!}} \det[\psi_{ij}(\boldsymbol{r}_k)]$. The modulus square of the wave function $|\Psi(\boldsymbol{r}_1,\ldots,\boldsymbol{r}_N)|^2$ is the probability density of finding the particles at positions $\boldsymbol{r}_1,\ldots,\boldsymbol{r}_N$.

In a single-shot measurement with a fluorescence microscope, a set of $N$ positions of atoms can be determined. It is therefore legitimate to study the outcomes of such measurements on a theoretical ground.  The positions are probabilistic variables, therefore the most probable ones are of special importance.  To determine the configuration maximizing the $N$-body probability distribution
$|\Psi(\boldsymbol{r}_1,\ldots,\boldsymbol{r}_N)|^2$ we used the Monte-Carlo algorithm \cite{Metropolis}. Starting from a randomly chosen configuration we shift positions of all particles and check if the
shifted configuration is more probable then the starting one. In case of failure another attempt is made.
\begin{figure}
\includegraphics[scale=0.29]{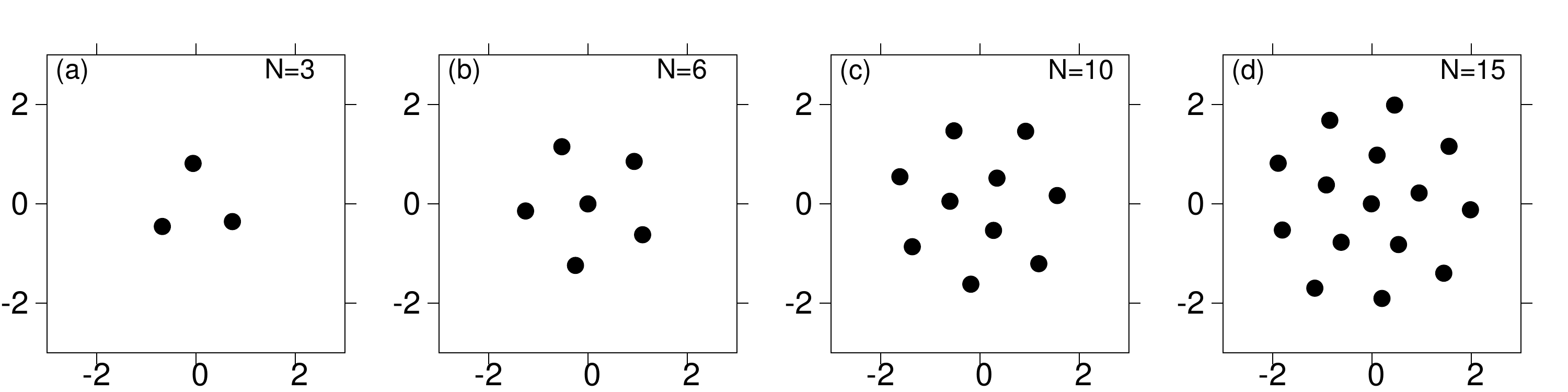}
\caption{{\bf Pauli crystals in two-dimensional harmonic trap.} Configurations maximizing $N$-particle probability: (a) -- 3 atoms, (b) -- 6 atoms,  (c) -- 10 atoms, (d) -- 15 atoms. }
\label{pauli}
\end{figure}
In  Fig.(\ref{pauli}) we show the most probable  configurations for a different number of fermions in a two-dimensional harmonic trap. We see that geometric structures do appear.

The patterns are universal if $N$ corresponds to closed energy shells, i.e. takes one of the values $N=1,3,6,10,15$. For open shells (not shown here) the patterns depend on the occupied orbitals at the Fermi level. Concentrating on the closed shells we see the following crystalline structures: an equilateral triangle for three atoms; a pentagon at the outer shell and one atom located at the trap center for six atoms; two shells are seen for ten atoms -- an equilateral triangle forming the inner shell and a heptagon forming the outer shell; and finally, for fifteen atoms, the third shell develops -- one atom is located at the center, five atoms at the middle shell form a pentagon and the remaining nine atoms are located at the outermost shell. Let us note that if the inner shell contains more than one atom it is generally not possible to match the discrete symmetries of the inner and outer shells. In this case the orientation of the inner shell with respect to the outer shell is fixed. Moreover the shells do not form regular polygons, i.e., distances of particles to the trap center vary slightly. The geometric shells are different than energy shells.

\section{Single shot detection of many-body system}
Existence of geometrical structures maximizing the N-body probability is an unexpected consequence of the Fermi-Dirac statistics. Whether this fact belongs to a class of  physical curiosities without any importance whatsoever depends upon possibility of detection of Pauli crystals. Do they really exist in a sense that the probability distribution of different configurations is sharply peaked at the most probable one? Or, on the contrary, are they very elusive object because probability distribution of different configurations is very flat and its maximum does not distinguish any particular geometric arrangement?

To answer these questions we have to analyze outcomes of single-shot measurements. Each such measurement gives a collection of values of $N$ particle positions. These values are unpredictable, have probabilistic character, however the most probable configurations should emerge as the most frequently observed ones in a series of measurements. Let us now discuss detection of particle positions, such measurement is particularly important in discussion of the properties of the many-body system.

Consider an array of detectors, each one  measures a particle at the position ${\bf X}$.  A single measurement of a particle at position $\bf x$ (a click in the measuring device) means that the detector reacted to a particle. We introduce a function that takes values 0 if no particle is detected and 1 if a particle is detected.:
\begin{equation}
{\rm Click}({\bf X}|{\bf x})=\delta(\bf X-\bf x).
\end{equation}
Because the outcome of a single measurement is unpredictable, one has to repeat it many times to get a statistics. Repeated measurements allow to make a histogram defined as:
\begin{equation}
h_M({\bf X})=\frac{1}{M}\sum^M_{s=1}{\rm Click}({\bf X}|{\bf x}^{(s)}),
\end{equation}
where $s$ refers to different measurements.
It can be shown straightforwardly that in the limit of infinitely many measurements one gets the one-particle
probability distribution:
\begin{equation}
\lim_{M\rightarrow\infty} h_M({\bf X})=p({\bf X}),
\end{equation}
where
\begin{equation}
p({\bf X})=\int d{\bf x}_2\cdots d{\bf x}_N |\Psi({\bf X},{\bf x}_2,\ldots,{\bf x}_N)|^2.
\end{equation}
This quantity gives the probability distribution of finding one particle at a point ${\bf X}$, without any information on the correlations between the particles.
\begin{figure}
\includegraphics[scale=0.6]{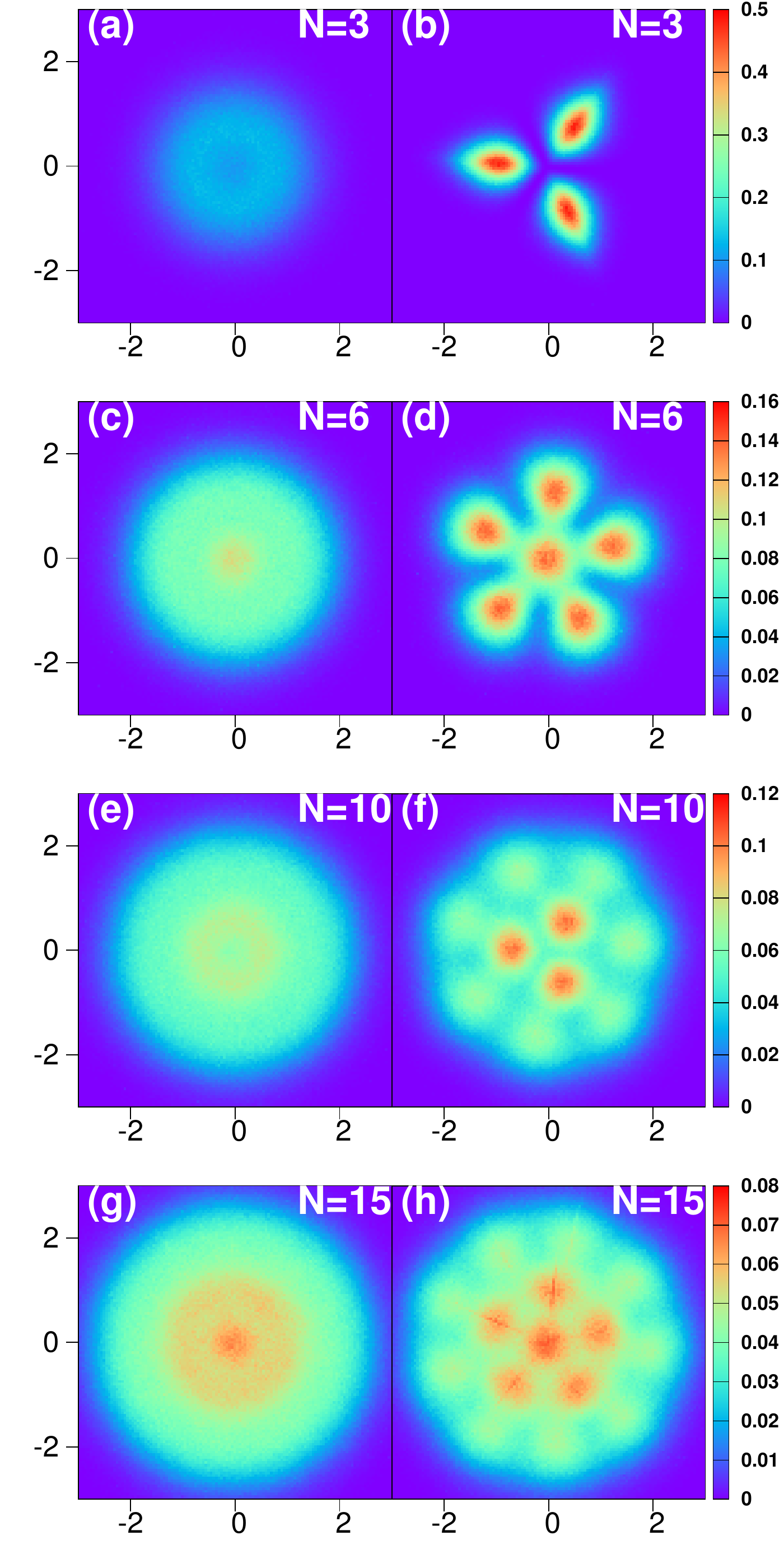}
\caption{{\bf Comparison of one-particle and configuration probability densities.} (a), (b) -- 3 atoms, (c), (d) -- 6 atoms, (e), (f) -- 10 atoms, (g), (h) -- 15 atoms. For each pair of figures we show a one-particle density distribution obtained with a direct collecting of the particle positions in many single shot experiments $H({\bf X})/N$ -- left panels: (a), (c), (e), (g). In right  panels -- (b), (d), (f), (h), we show configuration probability density $C({\bf X})/N$ resulting from the image processing. Position is measured in natural units of the harmonic oscillator.  The same color scale is used for every pair of figures. Note that configuration distributions are strongly peaked around maximal values. This maxima dominate over relatively flat structures of the one-particle density. }
\label{densities}
\end{figure}

Consider now a simultaneous detection of $N$ particles in a single shot measurement. Its result is given by:
\begin{equation}
{{\rm SingleShot}({\bf X}|{\bf x}_1,\ldots,{\bf x}_N}) =\sum^N_{i=1}{\rm Click}({\bf X}|{\bf x}_i).
\end{equation}
Single shot is, in our case,  a mapping of the $2N$-dimensional configuration space on the 2-dimensional 
physical space. It contains information on the geometry of the detected configuration, however it tells nothing about probabilities of different configurations. Many repetitions are needed to get the probabilities and to construct a histogram of particles' positions:
\begin{eqnarray}
H({\bf X})&=&\frac{1}{M} \sum^M_{s=1} {\rm SingleShot}({\bf X}|{\bf x}_1^{(s)} ,\ldots,
{\bf x}_N^{(s)})   \\
&=&\frac{1}{M}\sum_{s=1}^M\sum_{i=1}^N {\rm Click}({\bf X}|{\bf x}_i^{(s)}).
\label{nshot}
\end{eqnarray}
Evidently, by changing order of summation in Eq.(\ref{nshot}), we get:
\begin{equation}
H({\bf X})=N h_M({\bf X}).
\end{equation}
The histogram however, does not contain any information about higher order correlations, in particular about the geometry  carried by a single shot picture.  Correlations are washed out by summation of different outcomes.

\section{Correlating configurations}
Analysis of geometric configurations cannot be based on a simple histogram of particle positions.  Some quantitative methods allowing to  compare different  configurations, not the positions of individual particles, are required.  For a convenience we introduce a symbol $\{{\bf x}\}_N$  to denote the configuration $({\bf x}_1,\ldots,{\bf x}_N)$.  In order to compare an outcome of a measurement
$\{ {\bf x}\}_N $ with a given pattern, i.e. with the Pauli crystal structure  $\{ {\bf r}_{0}\}_N$, we have to define a measure in the space of configurations defining the distance between them. To this end we use  polar coordinates instead of the cartesian ones,
$({\bf x}_i)\rightarrow (r_{i},\phi_{i})$, $({\bf r}_{0_i})\rightarrow (r_{0_i},\phi_{0_i})$, and assign to  every
particle ${\bf x}_i$  its unique partner  ${\bf r}_{0_{\sigma(i)}}$, $({\bf x}_i)\rightarrow ({\bf r}_{0_{\sigma(i)}})$.
If the coordinates form a single shell then the transformation $\sigma$ is a cyclic permutation of the set
$1,\ldots,N$. We define the distance between the two configurations as:
\begin{equation}
d\left(\{{\bf x}\}_N,\{{\bf r_0}\}_N\right)=\sum^N_{i=1}\left(\phi_{0_i}-\phi_{{\sigma(i)}}\right)^2.
\end{equation}
The above definition is not the only possible. In fact a problem of the good definition of a distance between polygons  is one the basic problems in all pattern recognition algorithms which inevitably must assume some knowledge about the pattern. However, we checked that our definition works very well in the case studied here. We checked then when a system has a $n$-fold axis of symmetry, in order to ensure  
elementary fairness treatment of all particles, the maximal angle of rotation has to be limited to $2\pi/n$. Only then, all maxima of the pattern found have similar heights and widths. 

To observe the Pauli crystals one has to correlate outcomes of simultaneous measurement of all $N$ positions. Single shot will never give a pure geometry of the Pauli crystal because of quantum fluctuations of the particle positions. The crystalline pattern has to be extracted from the measured noisy structure with the help of the image processing. Our goal is to compare  different configurations leaving aside such details  as the position of the center of mass and the orientation of the configuration in space, thus the geometry of a configuration depends only on relative positions of particles. Therefore we shift the center of mass of the configuration
at hand to the origin of the coordinate system: ${\bf x}^{\prime}_i={\bf x}_i-{\bf x}_{CM}$
(${\bf x}_{CM}=(1/N)\sum_{i=1,N} {\bf x}_i$) and then apply rotations in the $x-y$ plane by an angle $\alpha$,
\begin{equation}
{\bf x}_i(\alpha)={\cal R}_{\alpha}\left({\bf x}_i-{\bf x}_{CM}\right).
\end{equation}
The `best alignment' of a given configuration $\{{\bf x}(\alpha)\}_N$ is therefore the one which minimizes the distance:
\begin{equation}
d\left( \{{\bf x}(\alpha) \}_N,\{{\bf r_0}\}_N\right)={\rm min.}.
\label{bestfit}
\end{equation}
Eq.(\ref{bestfit}) determines the rotation angle $\alpha$, which brings the given configuration to the `closest' distance to the pattern. Evidently this angle is different for every configuration.

Our strategy of image processing is the following. Each configuration, selected according to the $N$-particle probability distribution, is optimally transformed by an isometric transformation $\{{\bf x}\}_N \rightarrow \{{\bf x} (\alpha)\}_N$ to match the pattern according to Eq.(\ref{bestfit}). To gain an insight into the geometric configuration we introduce the configuration probability density , $C({\bf X})$ which is  {\it the histogram of configurations}:
\begin{eqnarray}
C({\bf X}) &=&\frac{1}{M} \sum^M_{s=1} {\rm SingleShot}({\bf X} | {\bf x}_{1}^{(s)} (\alpha),\ldots,
{\bf x}_{N}^{(s)}(\alpha))\nonumber \\.
\end{eqnarray}
The configuration probability density $C({\bf X})$ is seemingly not much different from the histogram
of particles' positions, $H({\bf X})$. In fact the difference, related to the preprocessing of the measurement outcome, is tremendous. Contrary to $H ({\bf X})$ which is proportional to one-particle probability density, the configuration probability density $C({\bf X})$ contains information about the geometric N-order correlations of the particles.

\section{Ensemble of configurations}
To generate an ensemble of configurations according to the many-body probability distribution we use the Metropolis algorithm. We generate a random Markov walk in the configuration space. The states belonging to the  Markov chain become members of the ensemble. The transition probability between subsequent configurations $\{ {\bf x}^{(s)} \}_N \rightarrow \{ {\bf y}^{(s)} \}_N$ is given by the ratio of their probabilities $p=|\Psi(\{{\bf y}^{(s)}\}_N)|^2/|\Psi(\{{\bf x}^{(s)}\}_N)|^2$, \cite{Metropolis}. If $p>1$ the trial  configuration is accepted to the chain:  $\{{\bf x}^{(s+1)}\}_N = \{ {\bf y}^{(s)}\}_N$. If $p<1$ there are two options chosen probabilistically: (a)  the trial step  is accepted  to the ensemble with the probability $p$,  $\{ {\bf x}^{(s+1)} \}_N=  \{{\bf y}^{(s)}\}_N$,  (b) the old configuration is again included into the chain with the probability $(1-p)$, $\{{\bf x}^{(s+1)}\}_N = \{ {\bf x}_N^{(s)}\}_N$. Typically we generate $2\times 10^6$ configurations, each being a set of $N$ positions on a two dimensional plane.
\begin{figure}
\includegraphics[scale=0.41]{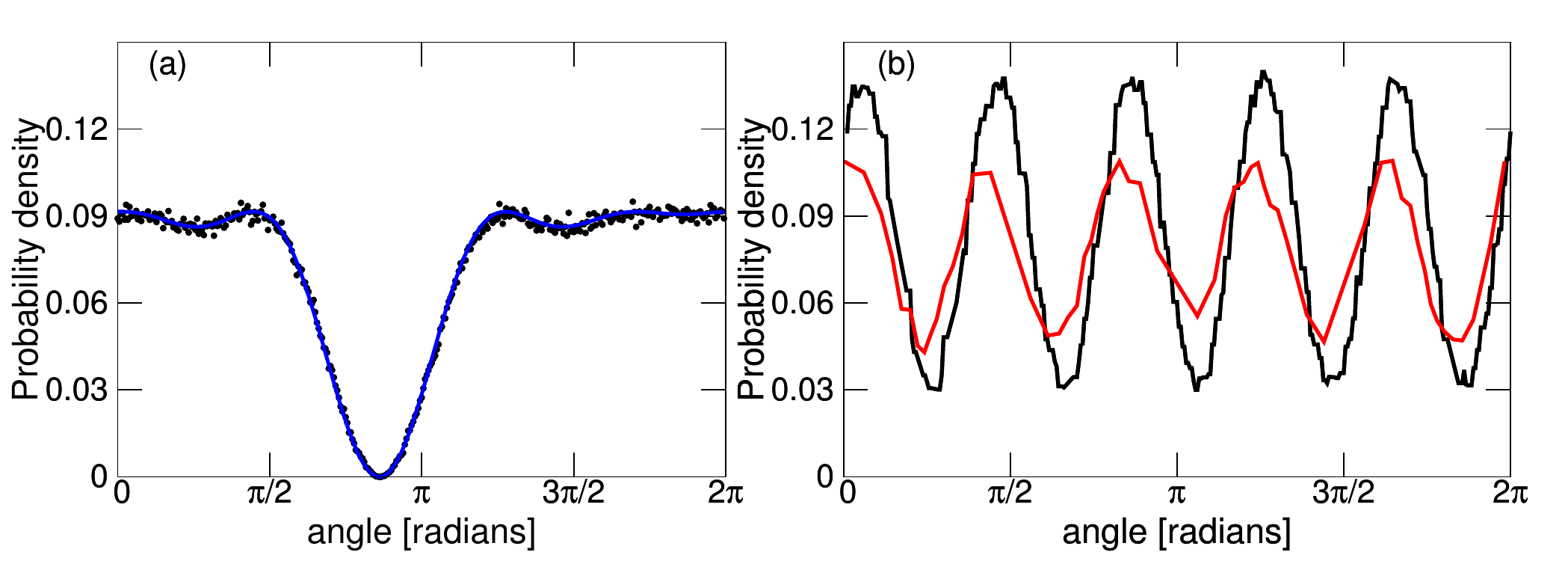}
\caption{{\bf Quality of pattern recognition.} (a) Configuration density of the excited state of  $6$-particle system
obtained after image processing based on a comparison with a {\it corresponding excited state} Pauli crystal pattern. (b) Configuration density of the state shown in (a) but obtained after processing of the same data as used in (a),  but based on a comparison with the {\it ground state} pattern of $6$-particle system. The patterns are marked by dots.}
\label{image_processing}
\end{figure}
Next we collect many realizations of the quantum state and after $M$ realizations we have $N\times M$ positions of particles. A histogram of such realizations, i.e. one-particle density, $H({\bf X})/N$, and configuration density probability, $C({\bf X})/N$,  for $N=3, 6, 10, 15$ atoms are shown in Fig.(\ref{densities}). In all cases the one-particle distribution is a smooth function of axial symmetry with some maxima in the radial direction. Clearly the one-particle distribution does not show any geometric structures resembling the Pauli crystals shown in Fig.(\ref{pauli}).

On the contrary, the configuration density probability $C({\bf X})/N$ shown in left panels of  Fig(\ref{densities}) exhibits the geometric structure of Pauli crystals. The agreement is amazing -- compare Fig.(\ref{pauli}). Quantum fluctuations lead to some smearing of the crystal vertexes, fortunately the uncertainties of atom positions are smaller than their separations, at least for small $N$.   For larger $N$ several shells are formed. The outer shells are somewhat melted because of quantum fluctuations.  A similar method of imaging geometrical structures formed by interacting Rydberg atoms was recently realized in experiment  with ultra cold atoms \cite{Bloch2}.
\begin{figure}
\includegraphics[scale=0.37]{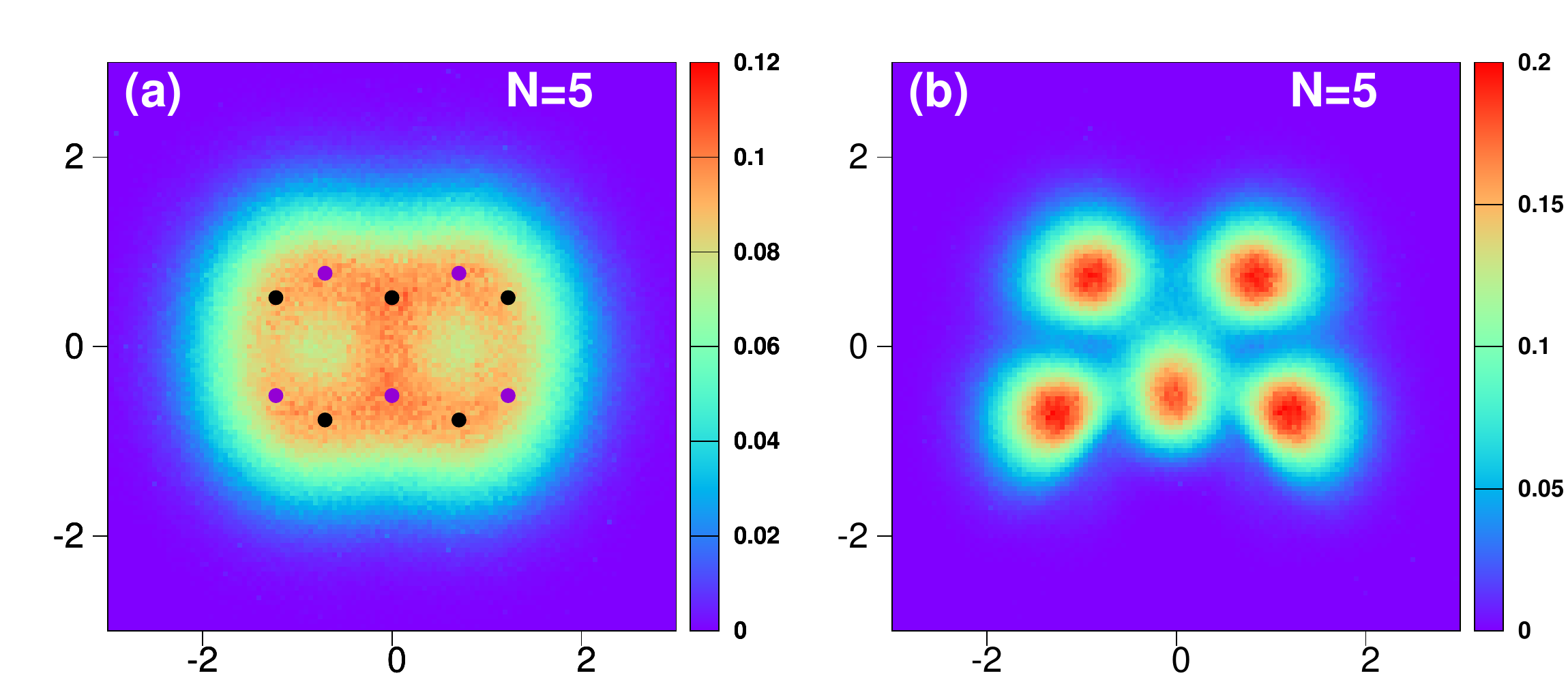}
\caption{{\bf Open shell Pauli crystalline structure for N=5 atoms.} (a) -- one-particle probability distribution $H({\bf X})/N$, (b) -- configuration probability distribution $C({\bf X})/N$. Note that color scale is different in both panels to emphasize a small structure in the one-particle density. Maxima of one-particle distribution do not coincide with maxima of configuration distribution. The latter are marked by blue and black dots. }
\label{oneparticle}
\end{figure}
Evidently our image processing, thus configuration density, $C({\bf X})$, depends on the pattern. To show how image-processed configurations  are biased by the pattern used, in Fig.(\ref{image_processing}) we show two configuration densities obtained by the best matching of the same ensemble of single shot pictures to a two different patterns. As an example we choose the  ensemble of configurations generated from the probability distribution of the one of lowest excited states  of $N=6$ particles, obtained by exciting the one at the Fermi surface.  In the Slater determinant we replaced the state  $n_x=2, n_y=0$ by $n_x=2, n_y=1$. In Fig.(\ref{image_processing}a) we show the configuration density obtained by fitting the ensemble of configurations to the 'native' crystalline structure of the excited state (marked by blue dots), while in the right panel, Fig.(\ref{image_processing}b), the same set of images is adjusted to the ground state Pauli crystal, marked by black dots.  A 'quality' of agreement, favors the native structure. If, as the pattern, a configuration similar to the native one were used, the pattern recognition algorithm would have produced a better agreement with the pattern . This however is not surprising, similar patterns are hard to distinguish.

In the case studied here the configuration of maximal probability is not unique. The system we investigate has some symmetries. The same symmetries are enjoyed by the $N$-particle probability. In the case of closed energy shells the symmetries are rotations around the trap center, reflections and inversion. There are also other symmetries like permutations of the particles and some specific symmetries depending on the particle number $N$. This results in a huge degeneracy of configurations with maximal probability. All of them differ by some symmetry operation. The symmetries are broken differently in each single realization. This is an additional reason why the histogram based on the generated single shot realizations washes out the Pauli-crystal structure. 

The above discussion might suggest that the problem of recognition of the crystalline structures is solely due the high symmetry of the system, and necessity of a proper alignment of single shot outcomes can be presumably overcame by choosing a trapping potential of a very low symmetry. One can hope then, that even one-particle density will show a number of maxima arranged in the geometry of Pauli crystals. Such small oscillations of one-particle density are in fact typical for small systems of noninteracting fermions as a result of the oscillatory character of one-particle wavefunctions -- thus of one-particle densities too. We want to stress that this is not the case here, structures we found are different.

To show the effect of symmetry, we consider a case of $N=5$ particles, i.e. the open shell structure where we have a freedom to choose two occupied orbital out of three basis states.  
In Fig.(\ref{oneparticle}) we show the one-particle density $H({\bf X})/N$ and the configuration probability density $C({\bf X})/N$  for the ground state system of $N=5$ particles. To lift the degeneracy we assumed that in the ground state the orbitals $n=2,m=0$, and $n=1,m=1$ are occupied and the orbital $n=0,m=2$ is empty. This choice is equivalent to assumption that $\omega_x$ is `a bit'  smaller than $\omega_y$. The ground state has no rotational symmetry, the only symmetry is the reflection with respect to the $y$-axis, $y \rightarrow -y$.

There are two equivalent configurations maximizing the $5$-particle probability. These are isosceles trapezoids differing by the reflection, see blue and black dots in Fig.(\ref{oneparticle}a). These Pauli crystalline structures are drawn on  top of the corresponding one-particle density. The structures are located in the region when the density is large, but evidently most of atoms forming the Pauli structure are not located at the maxima of the one-particle density. The one-particle density has two maxima, both on the $y$-axis. On the contrary, sharp maxima of the configuration density,  $C({\bf X})/N$, Fig.(\ref{oneparticle}b), fit perfectly to the geometry of the Pauli crystal. The configuration density was obtained by our image processing method using rotations to align the configurations. 

\section{Few-particle  correlations}
\begin{figure}
\includegraphics[scale=0.41]{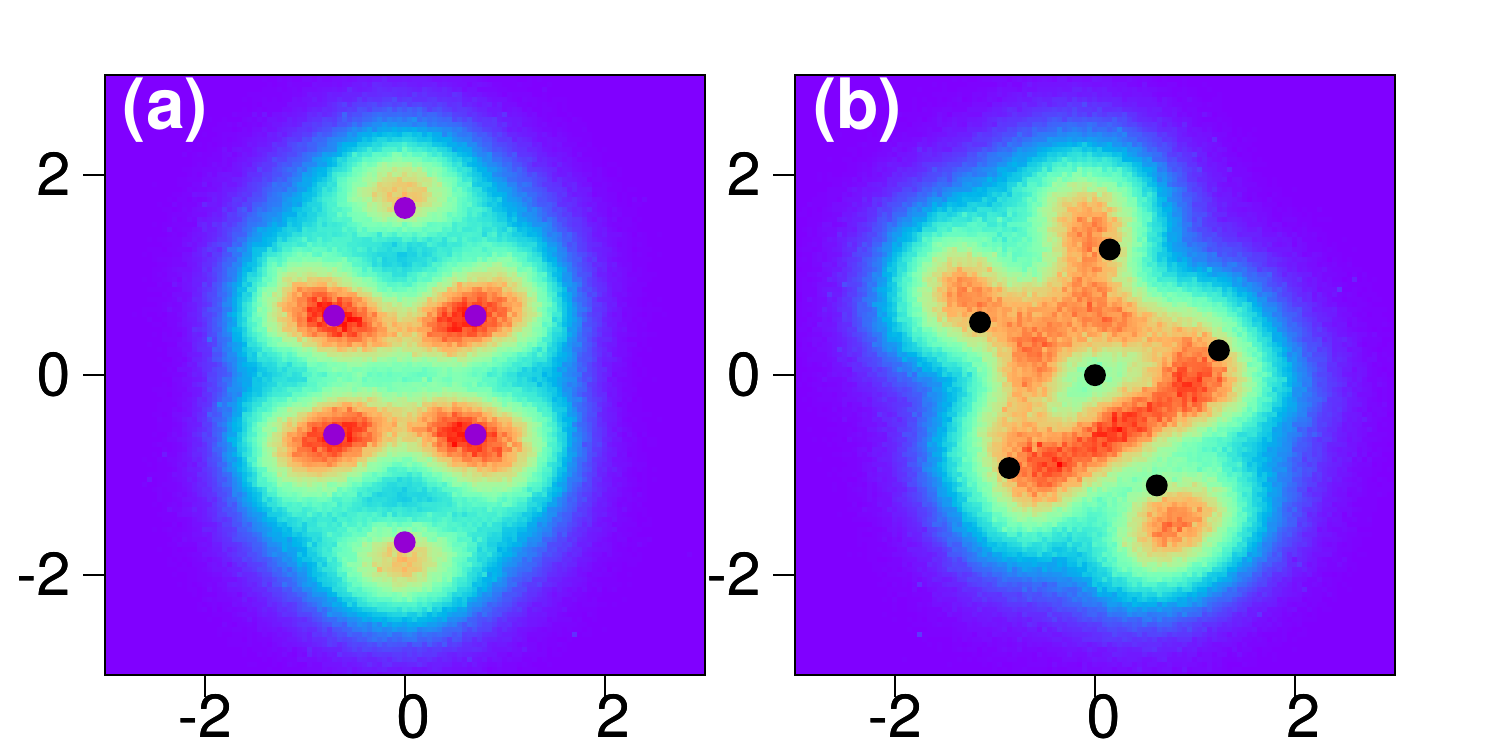}
\caption{{\bf Probability density distribution functions.} (a) - Two point correlation function - conditional probability density of detecting a particle at position $r_0=1.265$ (i.e. the radius of the Pauli crystal) as a function of the azimuthal angle, provided that another particle is found simultaneously at ($r_0$, $\phi_0=2.705$). Black scattered points result from the Monte Carlo simulations while the blue line is the exact analytic curve. Pauli blocking and kind  of the Friedel oscillations can be seen. These small four maxima in the correlation function indicate emerging Pauli crystal structure (b) - Configuration density as a function of the azimuthal angle at the distance  $r=r_0$ obtained from the histogram of configurations generated by the Markovian random walk after our image processing (black line). Five maxima corresponding to the vertexes of the Pauli crystal are clearly seen. Note high contrast. Red line - the same function plotted for a thermal state corresponding to $k_B T=\hbar \omega$.
Contrast is smaller.}
\label{correlations}
\end{figure}
In this section we use an example of  $N=6$  particles to show to what extend the low-order correlation function carry information on the Pauli crystalline structures.  The Pauli crystal in this case forms  two geometric shells with one particle in the trap center and five at the outer shell of the radius $r_0=1.265$, see Fig.(\ref{pauli}). The one-particle density does not depend on the azimuthal angle. This is expected because of the axial symmetry. But also a radial structure of the one-particle density does not indicate any geometrical arrangement  of atoms. The one-particle density has a sharp maximum at the center of the trap, a plateau at larger distances, and finally, at distance of the order of  $r \sim 1$,  it falls to zero quite rapidly, Fig({\ref{densities}c}). Nothing particular is happening at the distance $r_0=1.265$. The one-particle density does not suggest existence of the shell of the radius $r_0$.

One might expect, however, that two-body correlations will disclose a geometric ordering. Fig.(\ref{correlations}a)  shows the conditional probability density of particle detection at position $r_0$ as a function of the azimuthal angle, provided that simultaneously another particle is found at the same distance $r_0$  and at the azimuthal angle $\phi_0=2.705$. Polar coordinates $r_0$ and $\phi_0$ correspond to the location of  one of the vertices of the Pauli crystal in Fig.(\ref{pauli}). What is clearly seen is the effect of the Pauli exclusion principle (Pauli blocking) - the probability of finding the second particle close to the first one is very small. In addition weak oscillations are seen; they are of the same type as the Friedel oscillations \cite{Friedel} known in the case of electron gas.  No clear structure resembling pentagon is visible in Fig.(\ref{correlations}a),  however  four hardly distinguishable  maxima of the correlation functions are seeds of emerging structure.  The second order correlation function does not give enough evidence of existence of the Pauli crystal. In contrast, the image processing procedure described above, showing $N$-order correlations, unveils the crystalline structure. To support this statement we show in Fig.(\ref{correlations}b) a cut through the configuration density function $C({\bf X})$, Fig.(\ref{densities}d), along the circle of the radius $r_0=1.265$. Five distinct maxima indicate the most probable positions of particles arranged in a pentagon - the Pauli crystal. The contrast is very high.

\begin{figure}
\includegraphics[scale=0.38]{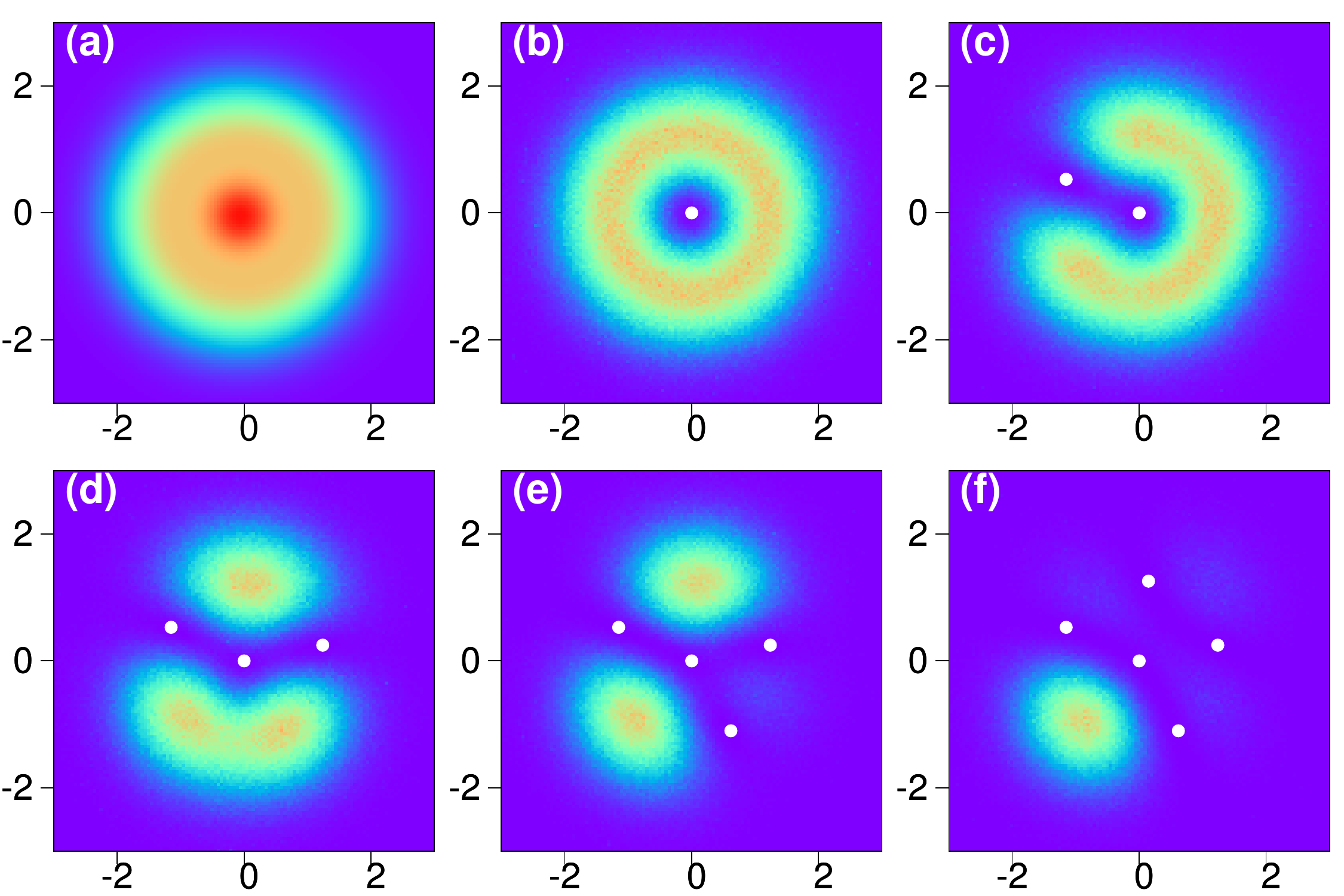}
\caption{{\bf Emergence of a geometric structures in a course of a conditional measurement.} 
Conditional density distributions of a ground state of a system of $N=6$ particles. Reference particles are marked by white dots. In every panel we show a higher order correlation function by adding a consecutive reference particle at the maximum of the preceding correlation function. All densities are normalized to the number of `not frozen' particles. (a)  One-particle density. (b) Conditional two-point probability of the same system - reference particle frozen at maximum of the function in (a), i.e. at ${\bf r}=0$. (c) Three-point correlation function -- two reference particles. (d) Four-point correlation function -- three reference particles. (e) Five-point correlation function -- four reference particles. (f) Six-point correlation function -- five reference particles. Note emergence of the Pauli crystalline structure. While consecutive particles are located in the vertices of the Pauli crystal, the corresponding conditional density distribution peaks more sharply around the positions of the remaining vertices of the structure.
}
\label{conditional}
\end{figure}
An alternative approach to the Pauli crystals is based on the method of Javanainen \cite{Javanainen}. In this approach the Pauli crystal should emerge from the hierarchy of the conditional probability functions. The starting point of this approach is to select a particle at position $\bf x_1$, then use the conditional probability to select the second particle at position $\bf x_2$, continue this way through three, four etc. conditional probabilities. One may expect that few particles will give hint on positions of all other particles. We verified this approach using example of 6 particles. In Fig.(\ref{conditional}) we show the result of this procedure. First, Fig.(\ref{conditional}a) we selected the first particle at the maximum of the one particle density. Corresponding one-particle conditional density 
shows a maximum along a ring of the radius of the Pauli crystal Fig.(\ref{conditional}b). This is the first signature of the emerging structure. Next we chose the position of the second particle on this ring. In Fig.(\ref{conditional}b)
we plot a corresponding three-point conditional probability.  Note a small structure appearing along the ring, Fig.(\ref{conditional}c), in addition to clearly visible Pauli hole.  When the third particle is chosen at the maximum on a ring, the Pauli structure of $N=6$ atoms system clearly emerges in higher order conditional distributions, Fig.(\ref{conditional}d)-Fig.(\ref{conditional}f).  The conditional approach to the high order correlation functions and emerging Pauli crystal structures is an independent test strengthening our confidence in the image processing method.

\section{Comparison with other systems and experimental prospects}
Many other systems exist that contain atoms or molecules arranged in a regular geometric structure, like molecules, crystals, clusters. Also more exotic structures can be formed, e.g. Wigner \cite{Wigner} and Coulomb crystals \cite{Mostowski,Diedrich,Wineland}. In the context of ultra cold trapped atoms interacting via a short range contact potential,  geometric crystalline structures - "Wigner molecules" were predicted \cite{Yannouleas,Baksmaty,Brandt}. In all these cases, however, the geometry is determined by a balance between attractive interactions at large distances and repulsive at small distances. Quantum statistics plays a marginal role in the resulting geometry in all cases. It should be stressed that the geometry of Pauli crystals differs on the fundamental level from that of other
crystals. It would be misleading to consider the anti-symmetry of the wave function as a simple kind of repulsion. The case of Pauli crystals is truly unique.
\begin{figure}
\includegraphics[scale=0.58]{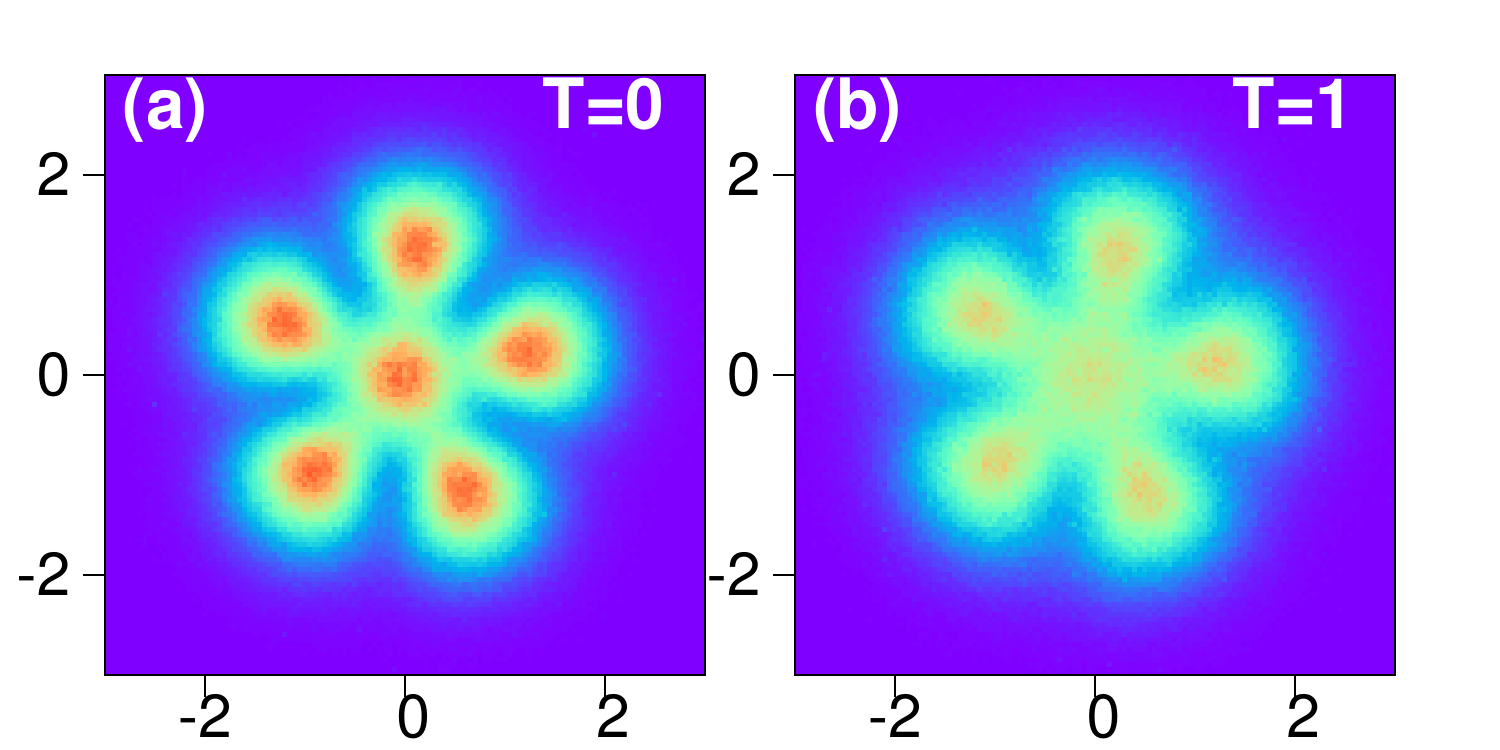}
\caption{{\bf Melting of the Pauli crystal at nonzero temperature.} (a)  Configuration distribution of the ground state of $N=6$ particle system.  (b) Configuration distribution of  the same system at nonzero temperature $T=\hbar \omega /k_B$.}
\label{melting}
\end{figure}
Observation of the Pauli crystals can be possible only in ideal or very weakly interacting quantum systems. Fermi-Dirac statistic leads to observable effects only when one-body wave functions of individual particles overlap. This is possible in the case of electrons in atoms. Electrons in atoms, however, are not good candidates for the envisaged experiments because of their Coulomb interactions.  We rather have in mind systems of ultra-cold fermion atoms in optical traps. Lithium $^{6}Li$ or Potassium $^{40}K$ atoms are good candidates. At densities of $10^{12}\,\mathrm{cm}^{-3}$ the wave functions describing atoms overlap at the temperature of the order of $T=10^{-7}\,\mathrm{K}$. These are the conditions at which quantum statistics plays a crucial role \cite{Anderson,Davis,DeMarco,Lewenstein}.

\section{Conclusions}
Our finding shows that even a simple system of noninteracting Fermi gas
has a geometry deeply hidden in many-body correlations. This finding might
suggest that geometric correlations are common in all Fermi systems.
Interactions compete with quantum statistics and modify the geometric
structures. For instance the Wigner crystals have different geometric
structures than the Pauli crystals. Therefore, one can think of systems
that will be somewhere between these two cases where both interactions and
statistics play a role in determining the geometric structure. This
suggests that the system may undergo some kind of 'geometric phase
transition' from one crystalline structure to another. We believe that
theoretical studies of high order geometric correlations in ultra cold
atomic systems, particularly in a view of experimental possibilities of
single shot pictures,  can bring to light many interesting and unexpected
information about the correlated many-body systems.

\section{Acknowledgments}
M.G. acknowledges support from the EU Horizon 2020-FET QUIC 641122. T.S. acknowledges financial support from the (Polish) Ministry of Science and Higher Education, Iuventus Plus 2015-2017 Grant ,,Strongly correlated systems of a few ultra-cold atoms'' (No. 0440/IP3/2015/73).
  \newcommand{\BibTitle}[1]{{\it ,,#1''},}
\newcommand{\BibName}[1]{}

\end{document}